# Gamma Ray Bursts Flares detected and observed by the Swift Satellite.

Guido Chincarini[1,2], Alberto Moretti[2], Patrizia Romano[1,2], Sergio Campana[2], Stefano Covino[2], Gianpiero Tagliaferri[2], Abraham D. Falcone[3], Davvid N. Burrows[3], Neil Gehrels[4], Vanessa Mangano[5], Giancarlo Cusumano[5], Matteo Perri[6], Paolo Giommi[6], Milvia Capalbi[6].

1) *Università degli Studi di Milano – Bicocca. Milano - Italy*
2) *INAF – Osservatorio Astronomico di Brera. Milano Italy.*
3) *Penn State University. State College, PA, USA*
4) *Goddard Space Flight Center. Greenbelt, MD, USA*
5) *INAF – IASF. Palermo, Italy*
6) *ASI Data Center. ROMA, Italy*

## ABSTRACT

The detection of flares with the Swift satellite triggered a lot of observational and theoretical interest in these phenomena. As a consequence a large analysis effort started within the community to characterize the phenomenon and at the same time a variety of theoretical speculations have been proposed to explain it. In this presentation we discuss part of the results we obtained analyzing a first statistical sample of GRBs observed with Swift. The first goal of this research is very simple: derive those observational properties that could distinguish between internal and external shock and between an ever active central engine and delayed shocks (refreshing) related to a very small initial Lorentz bulk factor. We discuss first the method of analysis and the morphology evidencing the similarities such flares have with the prompt emission pulses. We conclude that GRB flares are due to internal shocks and leave still open the question of whether or not the central engine is active for a time of the order of $10^5$ seconds after the prompt emission.

## 1. INTRODUCTION

The standard synchrotron shock model has been very successful in explaining the basic mechanism of the Gamma ray Burst (GRB) introducing the internal shock and external shock where the collision of ultra relativistic shells among themselves and with the circum stellar medium originate the observed prompt emission and the so called afterglow. To some extent and following a very abundant theoretical literature, Zhang and Meszaros (2004) and references therein this seemed to explain what we later will define as the X-ray canonical underlying light curve.

Following the detailed and dense sampling of the observations obtained with the X-ray Telescope on board of Swift, Gehrels et al. (2004), we firmly defined the characteristics of these light curves, Chincarini et al. (2005), Nousek et al. (2006), O'Brien et al. (2006). Such light curves, flux (or count rate) $\propto t^{-\alpha} \nu^{-\beta}$, show a very rapid decline during the first few hundred to a thousand seconds (with a spectral slope that with minor variations is about $\langle\beta\rangle \sim 1$) to flatten out or slightly invert their slope for the few following thousands of seconds to finally sharply decay with one or more break. The consensus seems that the first steep decay is mainly due to decay of the prompt emission while the causes of the following flattening are so far rather controversial. The two leading explanations: the injection of Energy as suggested by Panaitescu and Kumar (2000) and the off axis viewing observer's angle, Eichler and Granot (2006), are not completely satisfactory. The first calls for a rather uniform energy injection lasting for a long time since this part of the light curve is in general rather uniform, the second, which is certainly at work to some degree, may depend very strongly on the shape and uniformity of the pellets in the commoving frame of the shock. Sari (1997), on the other hand, sketches a rather interesting evolution for the afterglow that follows the prompt emission. With an initial Lorentz factor of the ejected material of a few hundreds the highly relativistic forward shock and the either Newtonian or relativistic reverse shock produces an increase in luminosity simply due to the increasing area of the shell with time. The combination of these two light curves, prompt emission and afterglow, may generate a canonical light curve of the type we observe by modulating the relative intensity and time evolution of the two components. The rather steep slope of the fading light curve does not agree however with the theoretically expected values. While the detailed physics related to the evolution of the luminosity it is not yet clear, this does not affect the present work where the canonical light curve is subtracted from the observations to isolate the superimposed flares as discussed in section 2.

The afterglow is generally referred to as due to the external shock; the collision of the forward shock with the Interstellar Medium (ISM). We rather refer to the circum-stellar medium (CSM). The CSM for a massive star (the probable progenitor of the GRB), eventually embedded in a recently formed HII region, is ionized to a distance larger than $10^{18}$ cm and fairly clean from dust since during the lifetime of the young massive progenitor star the radiation pressure cleaned up the surroundings. Most of these massive stars have a radiatively driven stellar wind with a velocity v ~ $(2\ G\ M\ /\ R)^{1/2}$ ~ 2 $10^8$ cm/s and a density running more or less as $r^{-2}$. The density at the radius (distance form the central engine) of the progenitor's photosphere may be much larger than the perused n = 1 particle $cm^{-3}$. We believe these are the most common characteristics of the CSM of a long GRB.

The X ray Telescope on board of Swift (XRT) monitors each detected object for a maximum time of about $10^6$ seconds (the flux of any GRB decreases fast as a function of time so that after this time is generally below the capability of the instrument), that is we cover a space in the source frame of at most R ~ $10^{18}$ - cm ~ 1 pc ( $R = \int_0^{10^6} 2\,\gamma^2(t)\, c\, \delta t$ ).

The forward shock may (or may not) encounter the unperturbed ISM only toward the end of our observing range. While other possibilities have been extensively discussed in the literature we believe this is the most realistic case for the CSM assuming the progenitor is a massive bright star. This is the stage in which the long GRB evolves. In addition we expect a cleaning action by the prompt emission jet. On the other hand we know about obscured and highly absorbed GRBs that may imply a more complicated environment at larger distances.

The short GRB (likely due to merging of NS-NS, NS – BH binaries) circum stellar environment is likely characterized in most cases by a very low density or absent ISM as expected in the outskirt of an early type galaxy or in an old low SFR galaxy (as the host of GRB050709). The progenitor itself does not produce much debris assuming it is a binary of relativistic stars.

In the pre-Swift era very few X-ray light curve of GRBs have been extensively sampled so that we could not have an accurate description of the light curve at these energies. In various cases, nevertheless, optical light curves detected variability (but for the X-ray see also Piro et al. 2005) superimposed to a very simple version of the canonical light curve described above and consistent with a smooth power law decay. None of the many explanations seemed to be satisfactory however and the limited observations could not allow a sensible diagnostic. Swift discovered early in the mission the presence of large flares superimposed to the underlying XRT light curve, GRB050406, Romano et al. (2006b), GRB050502B, Burrows et al. (2005).

Flares are finally detected in long and short, in low and high redshift GRBs. This perhaps is the most important and fundamental discovery attained by the Swift mission so far. It simply means that they are not directly related to the characteristics of the progenitor and to the particular mechanism by which the GRB is generated. It also means that the circum stellar material has a secondary role, if any, on the event and on the details of its evolution. In other words it means that the phenomenon is strictly related to the central engine and that quite likely the central engine generated subsequently to the long GRB event is similar to the central engine produced by the short burst event. This is a new window opened by Swift and since the onset of this research, still in progress; we realized that the flare activity could be a useful tool to understand some of the characteristics of the central engine.

In section 2 we will discuss the method of analysis and illustrates some of the flares detected in GRBs. Section 3 will deal with the flare morphology while section 4 will characterize the statistical sample. The results of the analysis will be described in section 5 while in the last section, section 6, we will conclude summarizing the main results and discussing the implication for the model.

In spite of the criticism expressed by Norris and Bonnel (2006) related to the classification of GRBs in long and short, and with which we fully agree, in this work for simplicity we will use this rather ill defined and instrument dependent classification. When needed we use a $H_0 = 70$, $\Omega_m = 0.3$ and $\Omega_\Lambda = 0.7$ cosmology.

## 2. ANALYSIS AND FITTING

Once the XRT data were converted in count rate we fitted the observed light curves using a multi-break (generally 2 and rarely 3) power law. On top of this "underlying canonical" light curve we added a Gaussian for each feature that could be reasonably identified as a bump or flare. In general we required at least 3 data points with an intensity of 2 -3 sigma above the canonical light curve to define a flare. One of the most common, and simple, light curve model that could be used has the following analytical expression:

$$k\left(\frac{t}{T_{braa}}\right)^{m}+\frac{a}{\left(\frac{t}{T_{brbb}}\right)^{n}+\left(\frac{t}{T_{brbb}}\right)^{p}}+b\,e^{-\frac{(x-x_0)}{2\,\sigma^2}}+....$$

where $T_{braa}$ and $T_{brbb}$ define the first and second light curve breaks. There is no physical interpretation of the GRB model attached to this analytical expression except that it characterizes the underlying light curve and the superimposed flare, however, in a very convenient way.

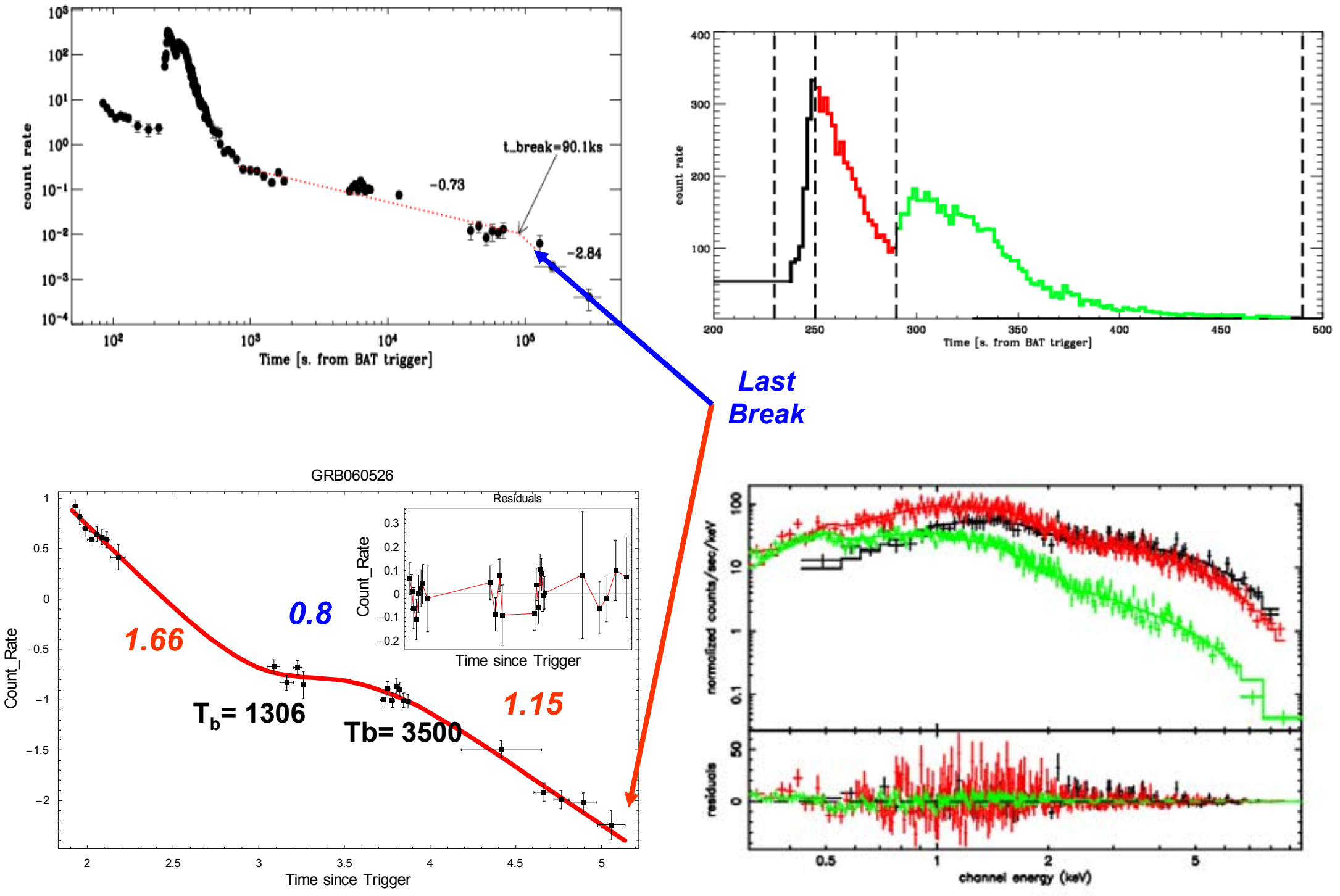

*Figure1. Top left: XRT data of GRB060526. Bottom left: Fit of the first $10^5$s of the underlying canonical light curve and residuals (inset). The slope and time breaks are marked in the plot. Top right: Details f the flare that clearly consists of two components. Bottom right: The three spectra, from harder to softer, cover respectively the rising curve of the first flare, the decay and the totality of the second flare (softer). The NH absorption and spectral index are respectively: {5 $10^{22}$, 1}, {3 $10^{22}$, 1.6}, {1 $10^{22}$, 2.5}.*

The fit of the underlying curve marks the slopes of the various phases and the change of slope (break). We decided to use the Gaussian curve for these “global” fits because of simplicity. Furthermore in this way we have, for most of the flare, a reasonable and unbiased measure of the width. The Gaussian does not accurately reflect the shape of the flare, however, as will be apparent from the morphology discussed in section 3.

Figure 1 illustrates with an example the procedure used for a recently observed GRB,. The evolution of the count rate as a function of time is given on the top left frame and it is clearly seen that a huge emission, flare, appears superimposed to a light curve that on a log log plane is very well approximated by a straight line, or by a broken straight line followed by an other steeper decay beginning at about $10^5$ s. Here the time, as always unless otherwise defined, is computed since the trigger time defined by the on board computer of the Burst Alert Telescope (BAT). Generally we carry out the first fit of the flare and underlying light curve simultaneously. However to better describe the process and illustrate a few details we plot on the bottom left of the figure the underlying light curve after elimination of the points characterizing the flare.

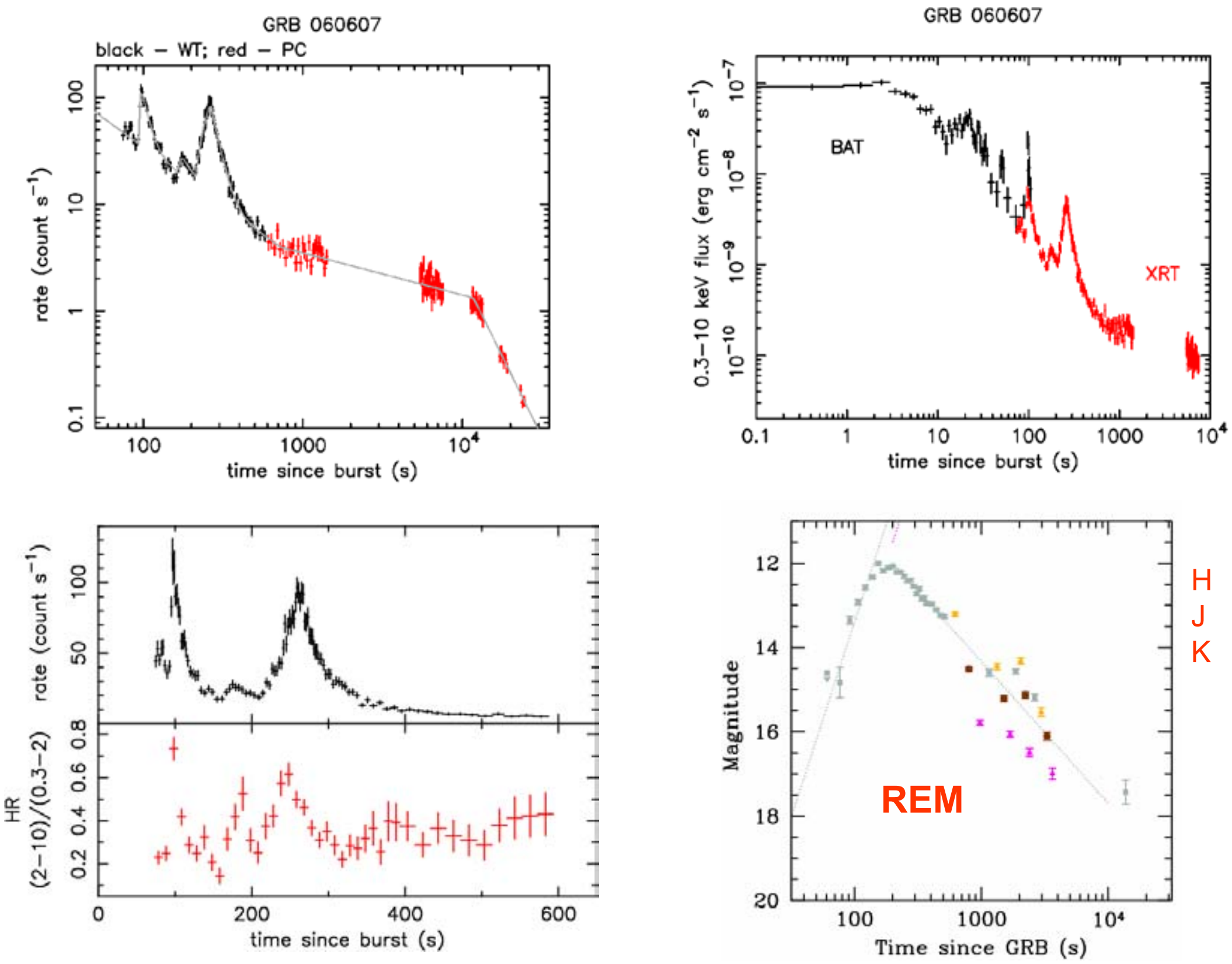

*Figure 2. From Top to Bottom: XRT observations of the GRB 060607 (two different observing modes), BAT and XRT observations of the first $10^4$ seconds, Rate and hardness ratio of the two main flares, near Infrared observations (H, J, K) of the very beginning of the flare obtained with the REM telescope at La Silla, Chile.*

To make sure that the parameters of the underlying curve are not affected by the data points of the flare we eliminated as many flare points as possible without sensible loss of statistics for the underlying curve. The light curve reflects clearly the characteristics of the standard canonical curve previously described. The isolated flare, a blend of two flares, is plotted on top right of the figure and the different colors (Astro-ph version) refer to the plot below (bottom right) where we have plotted the spectrum.
The rising part of the flare has a rather hard spectrum with a slope (β = 1) to be followed by a milder decay (β= 1.6). It is then followed by a rather soft flare (β=2.5).

These characteristics are quite common for flares. As a further example we plot in Figure 2 the time evolution of GRB060607. This burst beautifully shows the continuity between the BAT emission and XRT. The early flares have been observed both by BAT and XRT, see also GRB060124, Romano (2006a), and suggest indeed that the early flare activity is simply due to the long tail of the prompt emission as seen by XRT. This is by now a concept that receives consensus within the community and that however required a rather large set of very detailed observations. We should keep this possibility in mind when we model the central engine characteristics. The bottom left of Figure 2 shows simultaneously the flux evolution and the hardness ratio (H/S). It is impressive how the hardness ratio mocks the light curve and this has been observed in all of the flares with good statistics. The spectrum softens during the decay and hardens during the rising part. On the second flare, furthermore, there is an indication of a phase lag between the hardness ratio and the light curve. UVOT detected this GRB and monitored the rising to maximum and following decay. The bottom right plot shows the early light curve as obtained by REM, a robotic telescope that we installed at La Silla and works also in the near infrared [Covino et al. (2004), Chincarini et al. (2003), Zerbi et al.(2001)]. This is the first GRB maximum, prompt emission pulse, observed at all wavelengths (UVOT+REM = 170 to 2300 nm). Very similar properties, BAT+XRT, have been detected in the GRB060714.

Falcone et al. (2006) in the analysis of a statistical sample show that many flares have a non power-law spectrum which can be due to a hard spectrum or to a thermal component. Flare spectra, furthermore, look more like prompt emission spectra as opposed to afterglow spectra (the very bright flares seem however to have a somewhat softer spectral distribution when compared to the spectrum of the prompt emission). The analysis has been carried out on the sample we discuss below for the morphology and temporal characteristics and the conclusion is that in general a Band function is preferred. All observed spectra have been analyzed using a power law, Band function, cut-off power law, black body plus power law.

## 3. FLARE MORPHOLOGY

We find in the literature various descriptions on the light curve we should expect from the collision of relativistic shells with reference to the statistics and characteristics of the pulses observed during the prompt emission. Among these quite clarifying are the analysis by Kobayashi, Piran and Sari (1997) and Daigne & Mochkovitch (1998) among others. See for details also Dermer et al. (2004) and references therein. In brief if two masses, $m_f$ and $m_s$, move at different speed, fast and slow, with the slow preceding the fast shell, after collision part of the kinetic energy is transformed in internal energy, $E_{int}$ , and the forward shock proceeds with a Lorentz factor:

$$\gamma_{forward} = \sqrt{\frac{\dfrac{\gamma_s + 2\gamma_m}{\gamma_s}}{\dfrac{2\gamma_s + \gamma_m}{\gamma_s}}} \quad with \quad \gamma_m = \sqrt{\frac{\gamma_r \gamma_s (\gamma_r m_r + m_s \gamma_s)}{\gamma_s m_r + m_s \gamma_r}}$$

The peak of the emission being: $Luminosity_{Peak} = E_{Internal} / 2\,\gamma^2\,\delta t.$

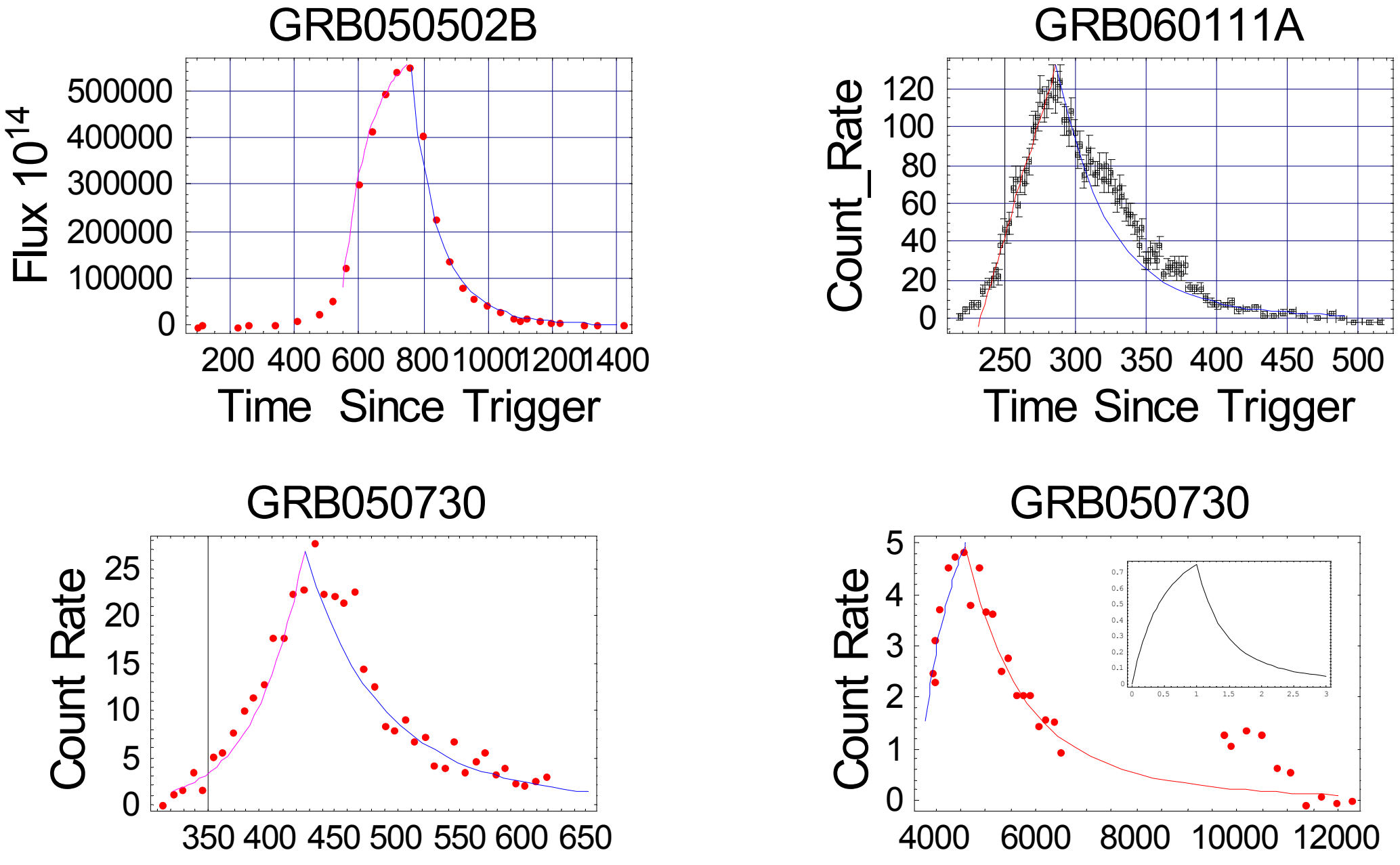

*Figure 3. From top left to bottom right: The very bright flare observed in GRB050502B showing an exponential rise and power law decay, the flare observed in GRB060111A where an excess of emission is present over the power law decay. A broken power law fits quite well the decay evidencing a few mini flares. This flare shows a morphology in which both rising and decaying curves are fitted by a power law. In the bottom right we show once more the classical flare as observed in GRB050502B. In the inset the theoretical profile of the emission expected from the collision of two relativistic shells.*

We are dealing essentially with a curve fast rising with an exponential law and decaying as a power law. Depending on the characteristics of the flare in the shock commoving frame we could have slightly different shapes in the observer frame as described by Shen et al. (2005).

In addition the power law decay of a flare must be constrained by the same rules governing the decay of any relativistic shock; in particular the exponent of the power law must not be larger than what is required by the curvature effect or naked burst decay, Kumar and Panaitescu (2000), Dermer (2004). The maximum slope is therefore dictated by the switching off of the source. Accounting for the radiation reaching the observer from the shell regions that are further away from the line of sight at the moment of the switching off , the temporal decay is related to spectral index by the relation: $\alpha = 2 + \beta$. This point will be further discussed in Section 5.

After isolating the flares from the underlying canonical light curve we obtain flares that have the morphology plotted in Figure 3. GRB050502B shows the expected classical form resulting from the shock of two colliding relativistic shells as plotted in the inset of the bottom right plot of Figure 3. This shape is seen in different GRBs. An other quite common morphology is that of a rising and decaying power law, bottom. As we can notice in the case of GRB050730, and in other GRBs, flares of different morphology are present. That is a given morphology is not related to a particular burst but, as expected, solely to the characteristics of the collision and of the shock. At the onset of the flare we often notice (always after subtraction of the underlying light curve) an excess of flux compared to the fit. This is within errors and likely due to the analysis procedure or to the presence of mini flares. These minor effects are hard to disentangle unless the GRB is very bright. In the case of GRB060111A, top right of Figure 3, we observe in the decaying part of the curve an excess of signal. This may be due to a blend of flares or to a particular decay of the light curve. Indeed in all those case where the count density rate is high enough we notice that flares are generally blended, have superimposed mini – flares and in some cases the decay mock the characteristics of the canonical underlying light curve as in the case of GRB060111A and GRB061117A, an effect that we call pseudo fractal. While we can not present an atlas in this context we try to illustrate the above with a few examples.

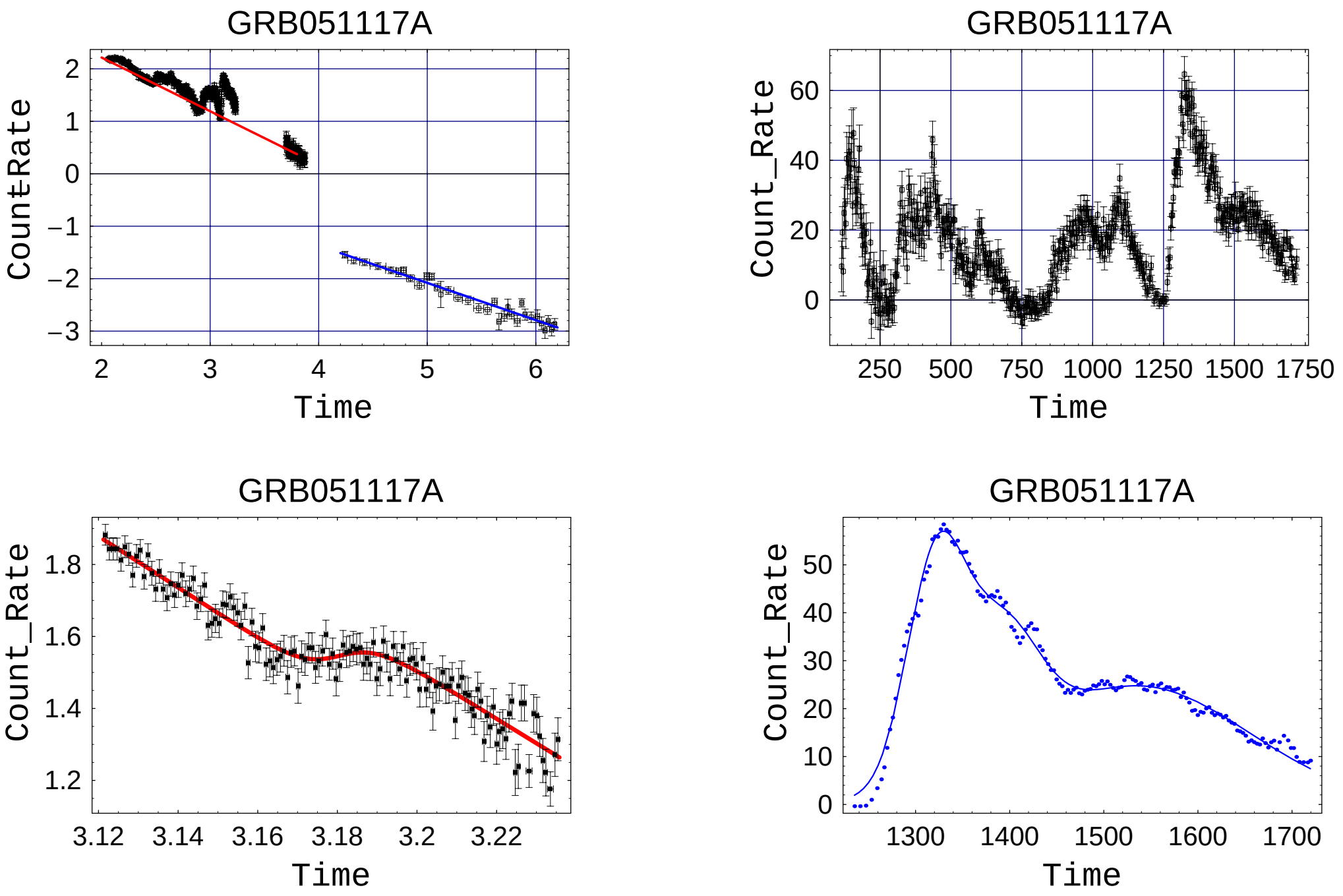

*Figure 4. Top left: XRT light curve. Top right: Flare light curve after subtraction of the underlying power law. Bottom left: the decay of the last flare fitted by a broken power*

*law. The parameters of the fit (note that we did not adjust $T_0$) are:*

$$52.98 \pm 0.61 \left( \frac{t}{1385.8 \pm 0.16} \right)^{7.11 \pm 0.44} + \frac{0.29 \pm 0.05}{\left( \frac{t}{1519.5 \pm 7.7} \right)^{-80 \pm 34} + \left( \frac{t}{1519.5 \pm 7.7} \right)^{6.3 \pm 1.9}} \quad with \quad \chi^2 = {}^{116}\!/_{124\ dof}$$

*Bottom right: Smoothing and fir of he last flare activity to evidence the mini flares.*

GRB051117A is a fantastic example to this end and for this GRB, where the density count rate is very high, Goad et al. (2006) discusses in detail the spectral evolution. Once we subtract the underlying light curve (straight line in the top left plot) and extract the variable part, plot top right of the figure, we witness a huge variability with a variety of blended flares. To single out flares is a difficult, and obviously somewhat subjective, task. Most interesting are the characteristics of the flare decaying curve starting at about 1320 s and ending at about 1750 s BAT trigger time, plot bottom left. Here we notice some mini variability superimposed to the decaying curve but its global characteristics can be fitted quite well using a broken power law as we normally do with the underlying canonical light curve. This is what we call pseudo – fractal characteristics. GRB051117A is not an isolated case. GRB060111A, Figure 3, and other GRBs show a similar characteristic. While this morphology does not imply directly a model we may tend to identify a mechanism for the canonical light curve that is similar to that acting in the decaying part of some flares. In this case we would exclude, for instance, geometry effects as suggested by Eichler and Granot (2006) and either refer to a particular characteristics of the decaying light curve to be explained or to a composite mechanism of the same nature. The model, however, must differ from the composite decay due to the prompt and afterglow emission, (see the model by Sari 1997). In this case (flares) we do not have the two components: prompt and afterglow emission. The variability is rather complicate and not fully understood yet.

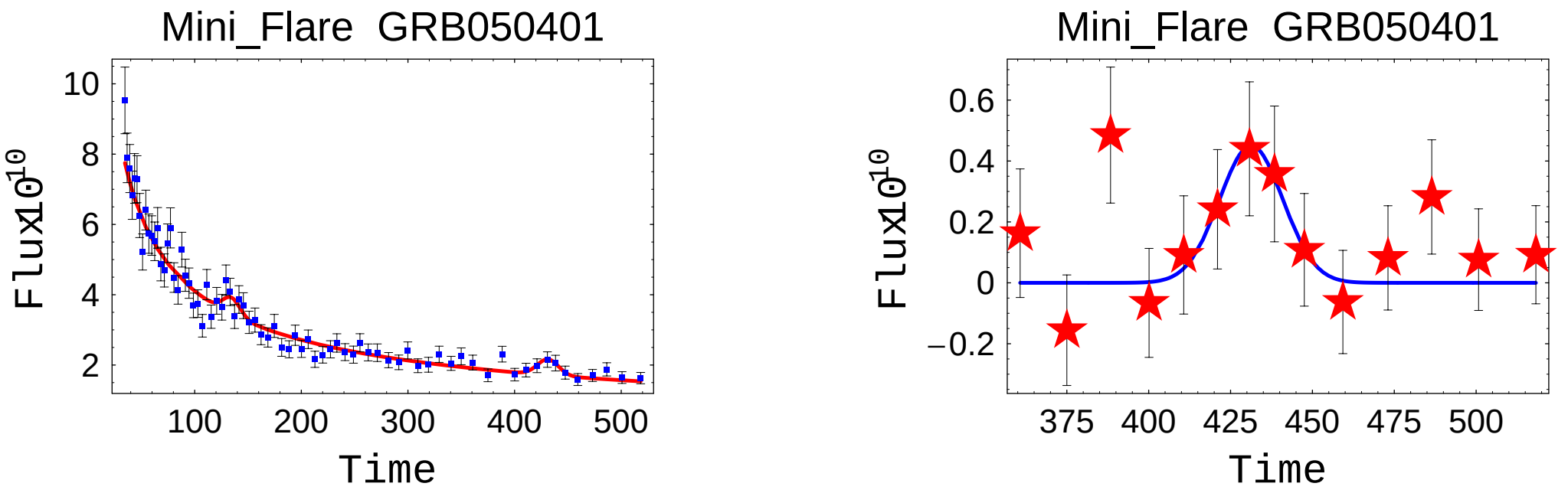

*Figure 5. The quite regular canonical light curve of GRB050401 shows nevertheless evidence of variability (mini-flares). On the right the mini flare after subtraction of the underlying power law light curve.*

To show the mini variability on the bottom right plot, Figure 4, we smoothed the last flare complex using a 5-point running mean. The continuous line is a fit done using 3 Gaussians. We can clearly see that superimposed to the fit, of course it could be done easily with a large number of Gaussians, we have wiggles that we identify as mini flares or mini variability. Such variability is present also in apparently normal canonical light curves. For instance GRB050401, see also Chincarini et al. (2005), is a quite regular

GRB with a canonical light curve starting off soon after the prompt emission with a mild slope. There is no evidence of flare activity. Nevertheless to a closer look, even if with low statistical significance, it is possible to detect very small flare as shown in Figure 5 where we used a linear scale to better evidence these little bumps.

## 4. THE STATISTICAL SAMPLE

The sample consists of all the GRBs observed by Swift up to the end of January 2006. For the analysis we selected 37 GRBs and measured 69 flares, see for details Chincarini et al. (2006), Falcone et al. (2006). A main concern was the presence of bias due to the sampling of the light curves since this depends strongly from the constraints of the satellite (period of the orbit, Earth and Moon zone of avoidance, pointing off axis of the Sun spacecraft axis direction and so on) and to the resolution of the observations. The resolution, or integration time per pointing, depends strongly from the count rate.

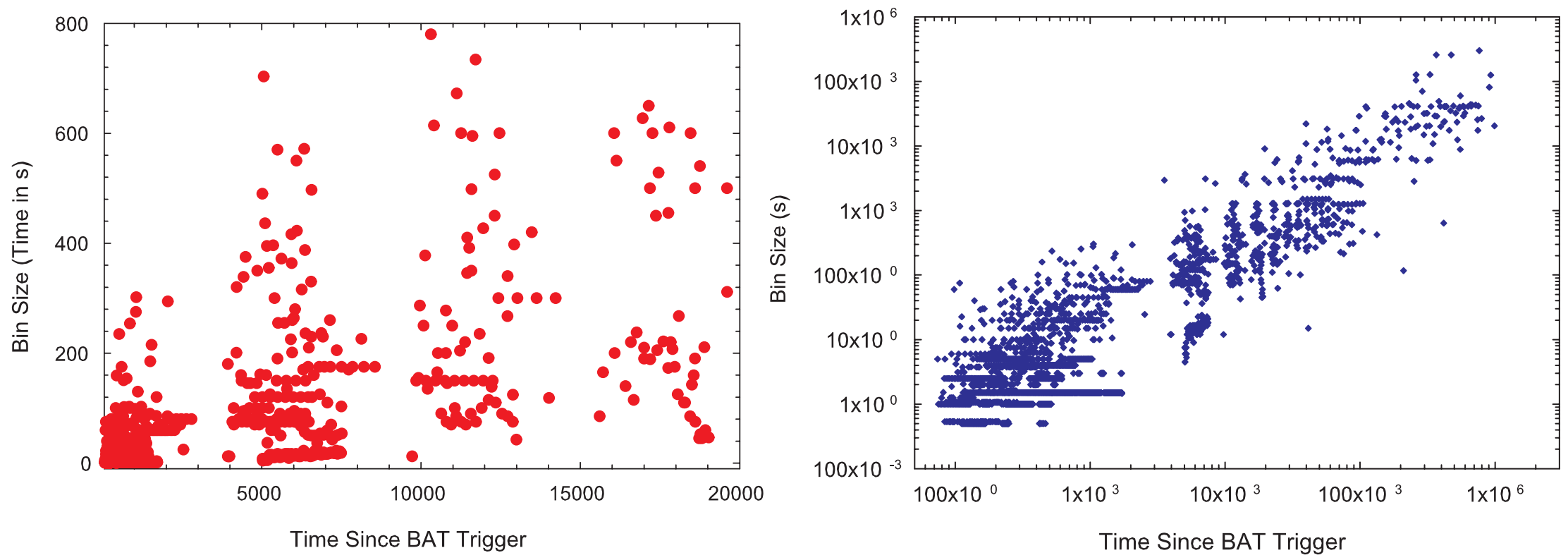

*Figure 6. Left: plot of the time bin for each data point as a function of BAT trigger time. The gaps are due to the orbit and pointing constraints. Right: as for the left plot in logarithmic coordinates and for the whole sample. The bin size increases with time due to the fading sources and the maximum integration tends to flatten out (maximum exposure time) at about 45000 seconds. Of course in some cases we made a much larger integration.*

In Figure 6 we plot the bin size in seconds for each data point we used. As it can be seen on the left plot the observations present gaps that are due to the various constraints and orbit of the satellite. Furthermore, as expected at the very beginning of the observations the resolution of the light curve is very high (Bin size small) since the count rate is rather high. To assess the presence of eventual bias and in particular to estimate whether or not we were missing high intensity flares toward the end of the light curve we made a large number of simulations. The results showed that we would have been able to detect narrow high intensity flares, similar to those detected in the early phase of the light curve, also toward the end of the light curve and that the observed distribution of flares was not biased by the way the observations were distributed as a function of time.

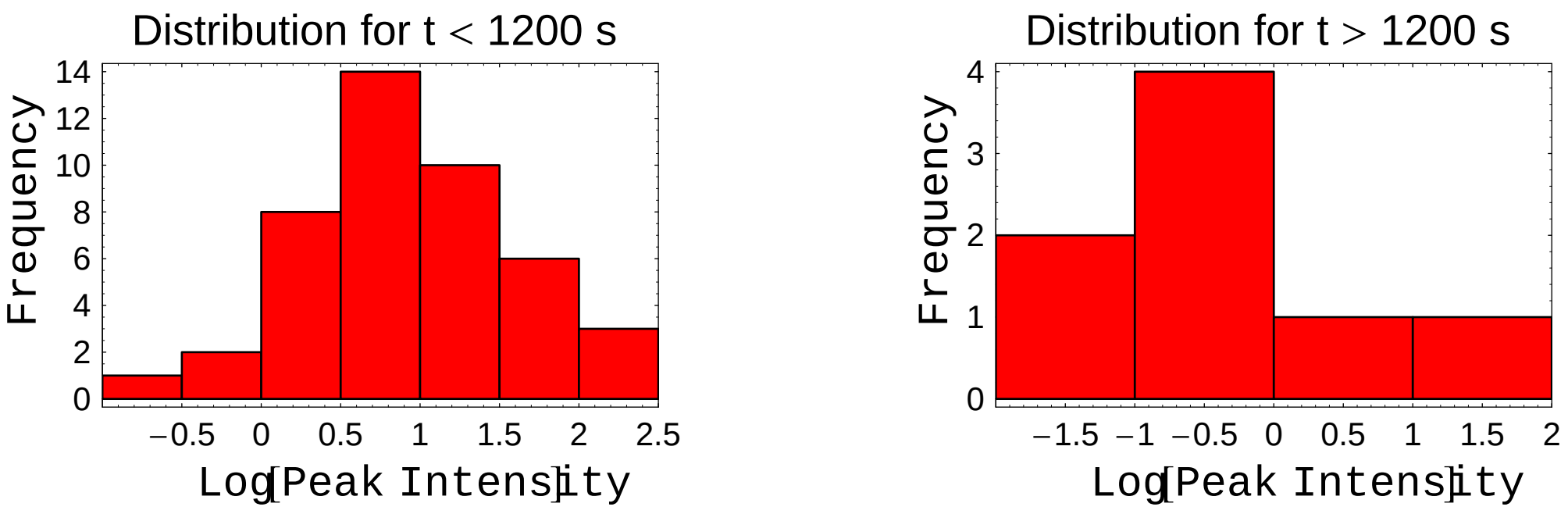

*Figure 7. Distribution of the Gaussian peak intensity of the flares observed within and after 1200 seconds.*

On the other hand we never detected a very intense flare late in the XRT light curve. The distribution of flare intensity we observe at time larger than 1200 s from the BAT trigger, Figure 7 right, peaks at an intensity of about 0.32 [counts $s^{-1}$] while the distribution of the flares that occur earlier than 1200 s peak at an intensity of 5.6 [counts $s^{-1}$], more than a factor 10 larger. Not only the two distributions are statistically different but we have indications that the intensity of the flares decreases linearly with time. We may have missed a few flares, but the simulations show that it is very unlikely that we would have missed flares with large intensity and small width. Therefore we must conclude that either we are dealing with collisions due to slow delayed shells or the central engine, if still active, lost the capability to produce short high intensity pulses.

## 5. RESULTS

Once a flare, unblended as far as we could tell, had been single out to mark its onset we estimated the time at which the flux in the light curve has a value of 1% the value of the flux at the peak of the flare ($T_{01}$). We then estimated the slope of the decay using the observed points between the peak and the end of the flare.

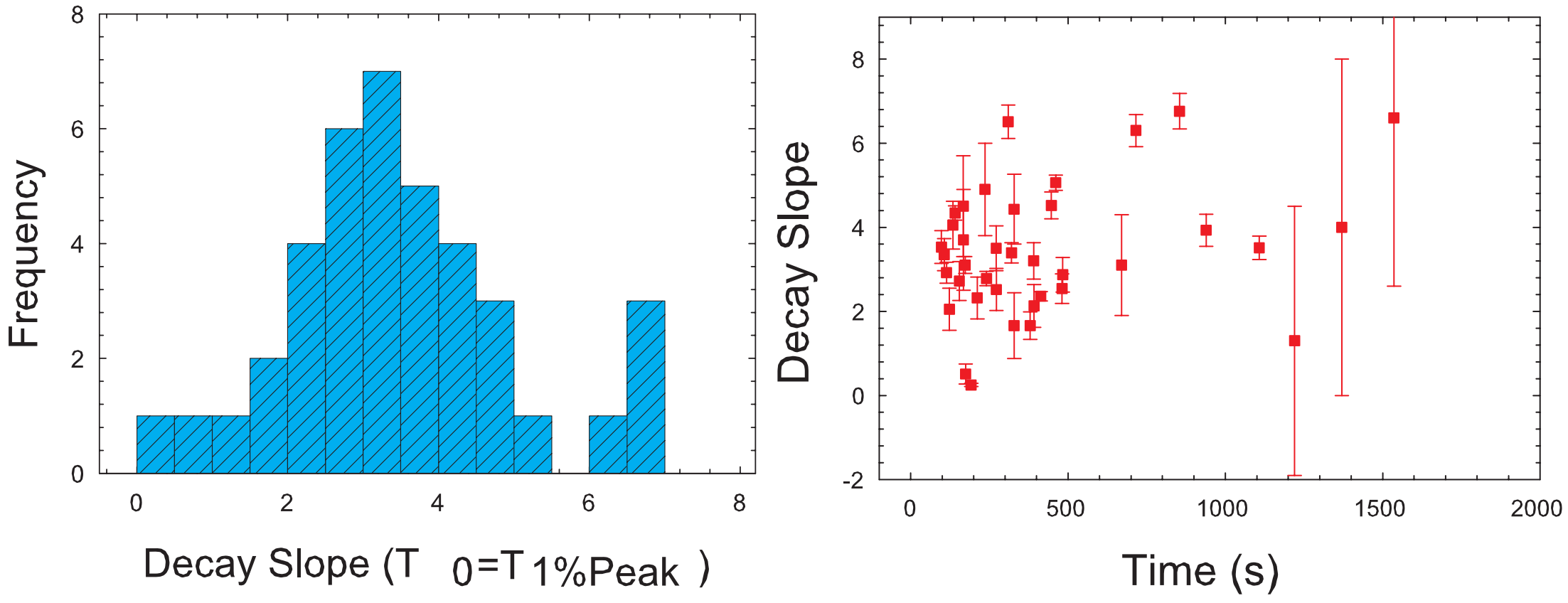

*Figure 8. Distribution of the decay slope of the total sample (left) and the decay slope as a function of time fort those detected after the first 2000 seconds after the BAT trigger.*

This definition is somewhat arbitrary and the slope we measured is sensitive to it. The curvature effect, 1.e. the maximum slope we can attain in a spherical shell by suddenly switching off the shell, may be reckoned also assuming a $T_0 = T_{1\%\ peak}$. This is a value better defined by the observations and all that is needed is to use the same reference time in the modeling. The slopes we determined in this way are displayed in the histogram, Figure 8 left. The plot shows that in some cases we measure rather high slopes in the decaying curve of the observed flares. Exceptionally high slopes, as in GRB050502B were also detected by Liang et al., 2006. In this GRB, however, the spectral index is rather high, $\beta \sim 2$, Falcone et al., 2006, and a high slope is expected. On the right of Figure 8 we have plotted the flare decay slope as a function of BAT trigger time; the mean value remains rather constant albeit a rather large spread.

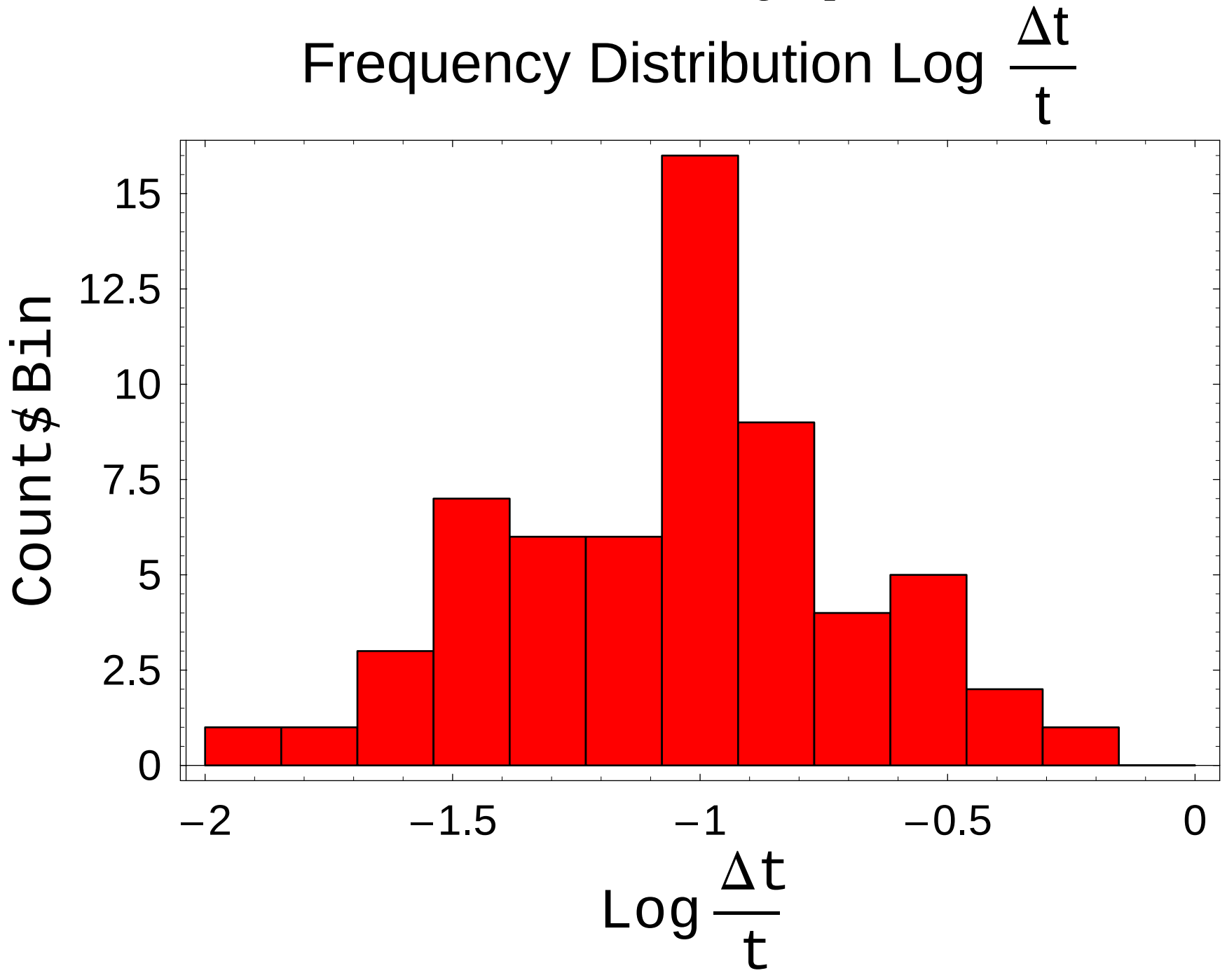

*Figure 9. Distribution of the ratio Flare's width (sigma of the Gaussian) – Peak time for the whole sample. For a sub-sample of flares with higher statistics the fit was done separately on the rising and falling curve using exponential and power law fits. In that case we derived the distribution of $T_{90}(flare)/T_{Peak}$ and obtain similar results, Chincarini et al.(2006).*

These values decrease if we use $T_0 = T_{5\%\ Peak}$ (as we did for a selected sub sample of well defined flares) or, obviously, in case we use $T_0 = T_{Peak}$. The use of the latter should finally be adopted for the ease and accuracy of the measure and to make it independent also from the rising time of the flare. In conclusion we the decay slope does not violate the naked burst paradigm.

In the proceedings of the Vulcano workshop, Chincarini (2006), we plotted the distribution of the ratio: Flare width (defined as the sigma of the Gaussian fit) / Time since the BAT trigger ($\Delta t/t$). The observations evidence: similarities to the prompt emission pulses, coincidence of the early flares with the tail of the prompt emission activity, Δt/t distribution. All of this in conjunction with the morphology and estimate of the bulk Lorentz factor (see also Falcone et al. 2006), helps in shedding light on the origin of the flares and consequently on whether or not the central engine remains active for a time of the order of at least 60000 seconds (last flare in GRB050724). In other words, independently from the detailed mechanism, we want to determine whether the central engine emits a number of pellets (shells) within a short time after the explosion with a rather broad Lorentz factor distribution (and likely different baryonic load) or it remains active for as long as needed. In the better case the late activity differs somewhat from the early activity since the characteristics of the flare are slightly different in spite of the fact that early and late flares are both very energetic.
On the other hand the shock collision of two shells resulting in two emissions of about the same internal energy lead to the relation $f_{Peak_1}\,\gamma_2^2 \big/ f_{Peak_2}\,\gamma_1^2 \propto \frac{\delta t_{e_2}}{\delta t_{e_1}}$ where $\delta t_e$ is the emission time scale, that is the time it takes to the reverse shock to cross the rapid shell, $f_{Peak}$ is the flux at the peak of the flare, Kobayashi et al. (1997). If $f_{Peak_1} \gg f_{Peak_2}$ *and* $\delta t_{e-2} \gg \delta t_{e-1}$. Referring to late ($f_{Peak_2}$ ,$\delta t_{e-2}$) and early ($f_{Peak_1}$ ,$\delta t_{e-1}$) flares we derive $f_{Peak_1} \big/ f_{Peak_2} \simeq 2200$ and $\frac{\delta t_{e_2}}{\delta t_{e_1}} \approx 1200$ so that we obtain for the ratio of the Lorentz factor: $\frac{\gamma_2}{\gamma_1} \approx 0.74$; the late collisions may have a Lorentz factor comparable to those occurring soon after the prompt emission. This may favor a central engine switching on again, At the time of writing Lazzati and Perna (2006) came out with good arguments supporting a late activity of the inner engine. We will further discuss this matter elsewhere.

## 6. SUMMARY AND CONCLUSIONS

Following a description of the circum stellar medium in which the GRB occur we show, using some examples to illustrate our statements, that the spectroscopic evolution of the flare goes from an hard spectrum during the rising to maximum to a softer one during the fading. The second flare in GRB060526 is softer than the first and this seems to be a characteristic of other flares. Such behavior is supported by plots of the hardness ratio as a function of time where it is possible to have also higher time resolution.

With this first sample of GRBs we are able to define a flare morphology that seems to be in good agreement with the expectations from collisions between clouds moving at relativistic speeds. We do not have yet a statistics of the type distribution since for this we need to analyze the complete large sample now at hand. We also mention in this work the existence of low intensity variability (mini flares).

The analysis of the flares in our sample show that the decaying slope of the flares is in agreement with the naked burst paradigm ($\alpha = 2 + \beta$) once we account for the reference time ($T_0$) and the spectral slope. For the first time we give the distribution of the width of the flare as a function of the BAT trigger time. We find furthermore that the peak intensity of the flares decreases as a function of BAT trigger time; late flares have a large width and rather low peak intensity with an energy (fluence) comparable to that of the early flares. These finding give important constraints on the mechanism at work, favor the internal shock and may lead to distinguish between an ever active central engine or, for late flares, the collision among low Lorentz factor shells (refreshing). That the results are not due to a bias in the sample has been demonstrated by extensive simulations. In conclusion we are getting close to defining the characteristics of the central engine as to guide the making of realistic models.

**Acknowledgments:**

This work is supported at OAB by ASI grant I/R/039/04, at Penn State by NASA contract NAS5-00136. We gratefully acknowledge the contributions of dozens of members of the XRT and BAT team at OAB, PSU, UL, GSFC, ASDC, and MSSL and our subcontractors, who helped make this instruments possible.

**References:**

Burrows, D.~N. Romano, P. Falcone, A., et al., Bright X-ray Flares in Gamma-Ray Burst Afterglows, Science 309, 1833-1835, (2005).

Chincarini G., Proceedings of the Vulcano Workshop 2006, Vulcano, Italy, May 22-27, Edited by F. Giovannelli & G. Mannocchi, ArXiv Astrophysics e-prints astro ph/0608414,(2006). .

Chincarini, G., Moretti, A., Romano, P., et al., 2005, Prompt and afterglow early X-ray phases in the comoving frame. Evidence for Universal properties?, ArXiv Astrophysics e-prints,astro-ph/0506453, (2005).

Chincarini, G., Zerbi, F., Antonelli, A., et al., The last born at La Silla: REM, The Rapid Eye Mount, The Messenger 113, 40-44, (2003).

Covino, S., Stefanon, M., Sciuto, G., et al., REM: a fully robotic telescope for GRB observations, Proceedings of the SPIE, 5492, 1613-1622, (2004).

Daigne, F., Mochkovitch, R., Gamma-ray bursts from internal shocks in a relativistic wind: temporal and spectral properties, MNRAS 296, 275-286, (1998).

Dermer, C. D., Curvature Effects in Gamma-Ray Burst Colliding Shells, ApJ 614, 284-292, (2004).

Eichler, D. and Granot, J., The Case for Anisotropic Afterglow Efficiency within Gamma-Ray Burst Jets, ApJL, 641, 5-8 (2006).

Falcone, A. D., Burrows, D. N., Lazzati, D., et al., The Giant X-Ray Flare of GRB 050502B: Evidence for Late-Time Internal Engine Activity, ApJ 641, 1010-1017 (2006).

Gehrels, N., Chincarini, G., Giommi, P., et al., The Swift Gamma-Ray Burst Mission, ApJ 611, 1005-1020 (2004).

Goad, M. R., The Swift Xrt Team, The prompt and early afterglow X-ray spectra of Swift GRBs. In: Holt, S. S., Gehrels, N., Nousek, J. A. (Eds.), American Institute of Physics Conference Series (2006).

Kobayashi, S., Piran, T., Sari, R., Can Internal Shocks Produce the Variability in Gamma-Ray Bursts? ApJ 490, 92 (1997).

Kumar, P. and Panaitescu, A., Afterglow Emission from Naked Gamma-Ray Bursts, ApJL, 541, 51-54 (2000).

Lazzati, D., Perna, R., X-ray flares and the duration of engine activity in gamma-ray bursts, ArXiv Astrophysics e-prints astro-ph/0610730 (2006).

Norris, J. P., Bonnell, J. T., Short Gamma-Ray Bursts with Extended Emission, ApJ 643, 266-275 (2006).

Nousek, J. A., Kouveliotou, C., Grupe, D., et al., Evidence for a Canonical Gamma-Ray Burst Afterglow Light Curve in the Swift XRT Data, ApJ 642, 389-400 (2006).

O'Brien, P. T., Willingale, R., Osborne, J., et al., 2006, The Early X-Ray Emission from GRBs,ApJ 647, 1213-1237 (2006)

Panaitescu, A. and Kumar, P., Analytic Light Curves of Gamma-Ray Burst Afterglows: Homogeneous versus Wind External Media, ApJ 543, 66-76 (2000).

Piro, De Pasquale, Soffitta, et al., Probing the Environment in Gamma-Ray Bursts: The Case of an X-Ray Precursor, Afterglow Late Onset, and Wind Versus ConstantDensity Profile in GRB 011121 and GRB 011211, ApJ, 623, 314 (2005).

Romano, P., Campana, S., Chincarini, G., et al., Panchromatic study of GRB 060124: from precursor to afterglow, A&A, 456, 917 (2006a).

Romano, P. Moretti, A. Banat, et al., X-ray flare in XRF 050406: evidence for prolonged engine activity, A&A 450, 59-68 (2006b).

Sari, R., Hydrodynamics of Gamma-Ray Burst Afterglow, ApJL 489, 37 (1997).

Shen, R.-F., Song, L.-M., Li, Z., Spectral lags and the energy dependence of pulse width in gamma-ray bursts: contributions from the relativistic curvature effect. MNRAS 362, 59-65 (2005)

Zhang, B. and Meszaros, P., Gamma Ray Bursts: properties, problems and prospects. International Journal of Modern Physics A, 19, 2385 – 2472 (2004).

Zerbi, R. M., Chincarini, G., Ghisellini, et al., The REM telescope: detecting the near infra-red counterparts of Gamma-Ray Bursts and the prompt behavior of their optical continuum, Astronomische Nachrichten, 322, 275-285, (2001).